\documentclass[prb,twocolumn,showpacs,preprintnumbers,amsmath,amssymb]{revtex4}
\usepackage{graphicx}

\begin{document}

\title{Asymmetry of the electronic states in hole- and electron-doped cuprates:\\
Exact diagonalization study of the $t$-$t'$-$t''$-$J$ model}

\author{T. Tohyama}
\email{tohyama@imr.tohoku.ac.jp}
\affiliation{Institute for Materials Research, Tohoku University, Sendai, 980-8577, Japan.}
\date{\today}

\begin{abstract}
We systematically examine the asymmetry of the electronic states in the hole- and electron-doped cuprates by using a $t$-$J$ model with the second-neighbor hopping $t'$ and third one $t''$ (the $t$-$t'$-$t''$-$J$ model).  Numerically exact diagonalization method is employed for a 20-site square lattice.  We impose twisted boundary conditions (BC) instead of standard periodic BC.  For static and dynamical correlation functions, averaging procedure over the twisted BC is used to reduce the finite-size effect.  We find that antiferromagnetic spin correlation remains strong in electron doping in contrast to the case of hole doping, being similar to the case of the periodic BC.  This leads to a remarkable electron-hole asymmetry in the dynamical spin structure factor and two-magnon Raman scattering.  By changing the twist, the single-particle spectral function is obtained for all momenta in the Brillouin zone.  Examining the spectral function in detail, we find a gap opening at around the $\mathbf{k}=(\pi,0)$ region for 10\% doping of holes (the carrier concentration $x=0.1$), leading to a Fermi arc that is consistent with experiments.  In electron doping, however, a gap opens at around $\mathbf{k}=(\pi/2,\pi/2)$ and persists up to $x=0.2$, being correlated with the strength of the antiferromagnetic correlation.  We find that the magnitude of the gaps is sensitive to $t'$ and $t''$.  A pseudogap is also seen in the optical conductivity for electron doping, and its magnitude is found to be the same as that in the spectral function.  We compare calculated quantities with corresponding experimental data, and discuss similarities and differences between them as well as their implications. 

\end{abstract}


\maketitle

\section{INTRODUCTION}

High-temperature superconductivity in cuprates emerges with carrier doping into insulating cuprates classified as the Mott insulator.  The carrier introduced into the cuprates is either an electron or a hole.  Although the symmetry of superconducting order parameter is common with $d$ wave in both cases,~\cite{Tsuei,Armitage1,Sato,Blumberg,Zheng,Snezhko} the phase diagrams exhibit asymmetric behaviors between the electron and hole carriers.  The most prominent difference appears in the antiferromagnetic (AF) region near the Mott insulator, where the AF order disappears with a small amount of carrier concentration ($x\sim 3$\%) in a hole-doped cuprate La$_{2-x}$Sr$_x$CuO$_4$ (LSCO), while in an electron-doped cuprate Nd$_{2-x}$Ce$_x$CuO$_4$ (NCCO) the AF order persists up to $x=0.15$.~\cite{Takagi}  A difference in magnetic properties is also seen in inelastic neutron scattering experiments: LSCO shows incommensurate spin structures for a wide range of $x$,~\cite{Yamada1} while in NCCO there is no incommensurate structures but  commensurate ones are observed.~\cite{Yamada2}  It is also an important difference that a spin-gap behavior observed in the underdoped region of hole-doped cuprates by the nuclear magnetic resonance experiments is not reported in electron-doped cuprates.~\cite{Zheng} 

Differences in the electronic properties between the hole- and electron-doped cuprates are also observed in other experiments.  The optical conductivity obtained from reflectivity measurements exhibits a pseudogap feature at around 0.2~eV in the AF phase of NCCO,~\cite{Onose} but there is no such a feature in LSCO with the same carrier concentration.  The occurrence of the pseudogap in the optical conductivity is correlated with the strong temperature dependence of the Hall coefficients and a metallic behavior in the $c$-axis resistivity.  From angle-resolved photoemission (ARPES) experiments, it is clearly observed that hole carriers doped into the parent Mott insulators first enter into the $\mathbf{k}=(\pm\pi/2,\pm\pi/2)$ points in the Brillouin zone and produce a Fermi arc,~\cite{Ino,Yoshida,Ronning} but electron carriers are accommodated at around $\mathbf{k}=(\pm\pi,0)$ and $(0,\pm\pi)$ and then the Fermi surface is formed in the superconducting region.~\cite{Armitage2}  The doping dependence of core-level photoemission also shows different behaviors of the chemical potential shift between NCCO and LSCO.~\cite{Harima} These experimental data indicate the difference of the electronic states between hole- and electron-doped cuprates.

The electronic states and magnetic properties in the electron-doped cuprates have been theoretically examined by many groups, focusing on the comparison with the hole-doped ones.~\cite{Tohyama1,Gooding,Tohyama2,Tohyama3,Lee,Yuan,Kontani,Kuroki,Kondoh,Manske,Yanase,Kobayashi,Kusuko,Markiewicz,Kyung1,Kyung2,Senechal,Kusunose}  Among them, the present author has studied a $t$-$J$ model with the second-neighbor hopping $t'$ and third one $t''$ (a $t$-$t'$-$t''$-$J$ model).~\cite{Tohyama1,Tohyama2,Tohyama3}  By applying the numerically exact diagonalization technique based on the Lanczos algorithm to small clusters, the dynamical spin structure factor, optical conductivity, single-particle spectral function, and thermodynamic properties have been calculated, and it has been pointed out that the difference of AF correlations caused by the presence of $t'$ and $t''$ is a prime source of the contrasting behaviors in the electronic states between the hole- and electron-doped cuprates.   

In small clusters used in our previous works,~\cite{Tohyama1,Tohyama2,Tohyama3} the momenta defined were discrete in the momentum space, since periodic boundary conditions (BC) were used.  Thus, momentum-dependent quantities such as the single-particle spectral function suffer from the discreteness.  Two-particle correlation functions also suffer from the finite-size effects under the periodic BC, because the two-particle operators are described as the convolution of the single-particle operators that are defined discretely in the momentum space.  Therefore, it is necessary to introduce a method that overcomes such discreteness and to clarify whether such a method changes the conclusions derived from small cluster calculations under the periodic BC.   In this paper, we introduce twisted BC for a 20-site square $t$-$t'$-$t''$-$J$ cluster.  The introduction of the twist can make the momenta defined continuous in the Brillouin zone, and thus we overcome the difficulty of the discreteness in the spectral function.~\cite{Tsutsui1}  For two-particle correlation functions, we introduce an averaging procedure over the twisted BC.  This procedure is known to reduce finite-size effects.~\cite{Poilblanc}  Therefore, results obtained under the twisted BC are expected to provide new information that has not been obtained under the periodic BC.  The quantities examined in this paper are, in addition to  the single-particle spectral function, several response functions in terms of spin and charge, i.e., the dynamical spin correlation, two-magnon Raman scattering, and the optical conductivity, together with several static correlations.

Being consistent with the previous works under periodic BC,~\cite{Tohyama1,Tohyama2,Tohyama3} we find that AF spin correlation remains strong in electron doping, in contrast to the case of hole doping, in the presence of the second- and third-neighbor hoppings $t'$ and $t''$.  This leads to a remarkable electron-hole asymmetry in the dynamical spin structure factor and two-magnon Raman scattering.  The single-particle spectral function also show dramatic differences between hole and electron dopings.  Along the nodal direction, i.e., the $\mathbf{k}=(0,0)$-$(\pi,\pi)$ direction, a quasiparticle band in the hole-doped system is gapless at the Fermi level as expected, while in electron doping the band is gapped up to the concentration $x=0.2$.  The gap is found to be correlated with the strength of AF correlation, indicating that the gap is magnetically driven.  The spectral function around the antinodal region, i.e., $\mathbf{k}=(\pi,0)$ region, shows contrasting behaviors: A gap appears in hole doping but not in electron doping, leading to a Fermi-arc behavior only in the hole-doped system.  The gap is found to be sensitive to $t'$ and $t''$, as is the case of the nodal gap in electron doping.  Such a Fermi arc behavior has not been detected under the periodic BC.  In the optical conductivity, a pseudogap clearly appears in the electron-doped system after the averaging procedure over the twisted BC.  The origin of the pseudogap is attributed to the strong AF correlation in the spin background.  The gap is found to have the same magnitude as that in the spectral function along the nodal direction, indicating the same origin.  In terms of pairing of carriers, we examine the $d$-wave pairing correlation function.   The pairing is found to be enhanced in the underdoped region of electron-doped system and also in the overdoped region of hole-doped one, being consistent with previous studies under the periodic and open BC.~\cite{White,Shih}  Since the quantities examined have fewer finite-size effects as compared with those under the periodic BC, we can make a more precise comparison between these results and experimental data.  From the comparison, we discuss similarities and differences between them as well as their implications.  

This paper is organized as follows.  In Sec.~\ref{Model}, we introduce the $t$-$t'$-$t''$-$J$ model and show outlines of the procedure to calculate the single-particle spectral function as well as the correlation functions for a 20-site square lattice under the twisted BC.  In Sec.~\ref{MagneticProperties}, calculated results of the doping dependence of magnetic properties such as the spin correlation functions and two-magnon Raman scattering are presented.  Being consistent with experiments, AF correlation remains strong in electron doping.  The single-particle spectral functions are shown in Sec.~\ref{SpectralFunction}.  Asymmetric electronic states between hole and electron dopings are discussed, focusing on gaps that appear in different momentum spaces.  Their implications are discussed compared with experimental data of ARPES.  In Sec.~\ref{ChargeDynamics}, the charge dynamics and pairing properties in the $t$-$t'$-$t''$-$J$ model are discussed.  The doping dependence of the optical conductivity clearly shows asymmetric electronic excitations that are closely related to the single-particle properties.  The $d$-wave pairing also shows remarkable asymmetric behaviors, and the role of the $\mathbf{k}=(\pi,0)$ states for the pairing is discussed.  The summary is given in Sec.~\ref{Summary}.

\section{Model and method}
\label{Model}

The Hamiltonian of a $t$-$J$ model with the second-neighbor hopping $t'$ and third one $t''$ (a $t$-$t'$-$t''$-$J$ model) reads 
\begin{eqnarray}
H&=& -t\sum_{\mathbf{i},\boldsymbol{\delta},\sigma}
    \left( \tilde{c}_{\mathbf{i}+\boldsymbol{\delta},\sigma }^\dagger \tilde{c}_{\mathbf{i},\sigma}+\tilde{c}_{\mathbf{i}-\boldsymbol{\delta},\sigma }^\dagger \tilde{c}_{\mathbf{i},\sigma} \right) \nonumber \\
&&  -t'\sum_{\mathbf{i},\boldsymbol{\delta}',\sigma}
   \left( \tilde{c}_{\mathbf{i}+\boldsymbol{\delta}',\sigma }^\dagger \tilde{c}_{\mathbf{i},\sigma }+\tilde{c}_{\mathbf{i}-\boldsymbol{\delta}',\sigma }^\dagger \tilde{c}_{\mathbf{i},\sigma } \right) \nonumber \\
&&  -t''\sum_{\mathbf{i},\boldsymbol{\delta}'',\sigma }
   \left( \tilde{c}_{\mathbf{i}+\boldsymbol{\delta}'',\sigma }^\dagger \tilde{c}_{\mathbf{i},\sigma }+\tilde{c}_{\mathbf{i}-\boldsymbol{\delta}'',\sigma }^\dagger \tilde{c}_{\mathbf{i},\sigma } \right)  \nonumber \\
&&  +J\sum_{\mathbf{i},\boldsymbol{\delta}}
      \mathbf{S}_{\mathbf{i}+\boldsymbol{\delta}}\cdot \mathbf{S}_\mathbf{i} \;,
\label{H}
\end{eqnarray}
with $\boldsymbol{\delta}=\mathbf{x}$ and $\mathbf{y}$, $\boldsymbol{\delta}'=\mathbf{x}+\mathbf{y}$ and $\mathbf{x}-\mathbf{y}$, and $\boldsymbol{\delta}''=2\mathbf{x}$ and $2\mathbf{y}$, $\mathbf{x}$ and $\mathbf{y}$ being the unit vectors in the $x$ and $y$ directions, respectively.  The operator $\tilde{c}_{\mathbf{i},\sigma}=c_{\mathbf{i},\sigma}(1-n_{\mathbf{i},-\sigma})$ annihilates a localized particle with spin $\sigma$ at site $\mathbf{i}$ with the constraint of no double occupancy, and $\mathbf{S}_\mathbf{i}$ is the spin operator at site $\mathbf{i}$.  In the model, the difference between hole and electron doping is taken into account by the sign difference of the hopping parameters together with the difference of the localized particle.~\cite{Tohyama1} For hole doping, the particle is an electron with $t>0$, $t'<0$, and $t''>0$, while the particle is a hole with $t<0$, $t'>0$, and $t''<0$ for electron doping.  Although the ratios $t'/t$ and $t''/t$ are material dependent,~\cite{Tohyama4} we take in this study $t'/t=-0.25$ and $t''/t=0.12$ for both the hole- and electron-doped cases, which is obtained by fitting the ARPES Fermi surface for NCCO with $x=0.15$.~\cite{Armitage3}  The $|t|$ is usually taken to be $~0.35$~eV.~\cite{Tohyama4}  We set $J/|t|=0.4$.  Hereafter, $\hbar=e=1$, and the distance between the nearest-neighbor sites in the two-dimensional lattice is set to be unity.

In order to examine the single-particle spectral function in the model, we use the exact diagonalization method for an $N$-site square lattice with the translational vectors $\mathbf{R}_a=l\mathbf{x}+m\mathbf{y}$ and $\mathbf{R}_b=-m\mathbf{x}+l\mathbf{y}$, being that $l,m\geqslant 0$ and $N=l^2+m^2$.  If periodic BC are used for the lattice, the momentum $\mathbf{k}_0$ for single particle is defined as $\mathbf{k}_0=2\pi (l n_1- m n_2)/N\mathbf{x}+2\pi (m n_1-l n_2)/N\mathbf{y}$, $n_1$ and $n_2$ being integers that give discrete $N$ points in the first Brillouin zone.  Introducing BC with twist, we can define momenta continuously in the Brillouin zone.~\cite{Tsutsui1} This procedure gives smooth band structures even if we use finite-size lattices.  The twist induces the condition that $\tilde{c}_{\mathbf{i}+\mathbf{R}_a,\sigma}=e^{\mathrm{i}\phi_a}\tilde{c}_{\mathbf{i},\sigma}$ and $\tilde{c}_{\mathbf{i}+\mathbf{R}_b,\sigma}=e^{\mathrm{i}\phi_b}\tilde{c}_{\mathbf{i},\sigma}$, with arbitrary phases $\phi_a$ and $\phi_b$.  Note that $\phi_a=\phi_b=0$ ($\pi$) corresponds to the periodic (antiperiodic) BC.  Introducing an arbitrary  momentum $\boldsymbol{\kappa}=\kappa_x\mathbf{x}+\kappa_y\mathbf{y}$ defining $\phi_{a(b)}=\boldsymbol{\kappa}\cdot\mathbf{R}_{a(b)}$, the momentum for a given $\boldsymbol{\kappa}$ reads 
\begin{equation}
\mathbf{k}=\mathbf{k}_0+\boldsymbol{\kappa}\;.
\label{k}
\end{equation}
In order for $\mathbf{k}$ to cover the full Brillouin zone, $\boldsymbol{\kappa}$ needs to scan a square with the four corners that $(\kappa_x,\kappa_y)=\pm\pi/N(l-m,l+m)$ and $\pm\pi/N(l+m,-l+m)$.  We note that imposing the twist is equivalent to transforming the operator $\tilde{c}^\dagger_{\mathbf{i},\sigma}\tilde{c}_{\mathbf{j},\sigma}$ in (\ref{H}) into $e^{\mathrm{i}\boldsymbol{\kappa}\cdot(\mathbf{R}_\mathbf{j}-\mathbf{R}_\mathbf{i})}\tilde{c}^\dagger_{\mathbf{i},\sigma}\tilde{c}_{\mathbf{j},\sigma}$, $\mathbf{R}_\mathbf{i}$ being the position of site $\mathbf{i}$.  The twist changes the hopping terms but not the exchange term in (\ref{H}). 

For a given $\boldsymbol{\kappa}$, the single-particle spectral function $A(\mathbf{k},\omega)$ at zero temperature reads
\begin{equation}
A(\mathbf{k},\omega)=A_-(\mathbf{k},\omega)+A_+(\mathbf{k},\omega)\;,
\label{Akw}
\end{equation}
with
\begin{equation}
A_\pm(\mathbf{k},\omega)=\sum_{m,\sigma} \left|\left< \Psi^{\boldsymbol{\kappa}}_m \left| a_{\mathbf{k},\sigma} \right|\Psi^{\boldsymbol{\kappa}}_0\right>\right|^2 \delta(\omega\mp (E^{\boldsymbol{\kappa}}_m - E^{\boldsymbol{\kappa}}_0)-\mu_{\boldsymbol{\kappa}})\;,
\label{Akw+-}
\end{equation}
where $A_+$ ($A_-$) is the  electron-addition (electron-removal) spectral function.  For hole doping, $a_{\mathbf{k},\sigma}=\tilde{c}_{\mathbf{k},\sigma}^\dagger$ and $\tilde{c}_{\mathbf{k},\sigma}$ for $A_+$ and $A_-$, respectively, $\tilde{c}_{\mathbf{k},\sigma}^\dagger$ ($\tilde{c}_{\mathbf{k},\sigma}$) being the Fourier component of $\tilde{c}_{\mathbf{i},\sigma}^\dagger$ ($\tilde{c}_{\mathbf{i},\sigma}$) with momentum $\mathbf{k}$ defined in (\ref{k}) and spin $\sigma$.  On the other hand, $a_{\mathbf{k},\sigma}=\tilde{c}_{\mathbf{k},\sigma}$ and $\tilde{c}_{\mathbf{k},\sigma}^\dagger$ for $A_+$ and $A_-$, respectively, for electron doping.  The $\Psi^{\boldsymbol{\kappa}}_0$ and $\Psi^{\boldsymbol{\kappa}}_m$ represent the ground state with the energy $E^{\boldsymbol{\kappa}}_0$ and the final state with $E^{\boldsymbol{\kappa}}_m$, respectively, for a given $\boldsymbol{\kappa}$.  The chemical potential $\mu_{\boldsymbol{\kappa}}$ is also dependent on $\boldsymbol{\kappa}$, which is defined as one half of the energy difference between the first ionization and affinity states of the system.  In this study, we calculate $A(\mathbf{k},\omega)$ for a lattice with $N=20$ $(l=4, m=2)$ using a standard Lanczos technique with a Lorentzian broadening of $0.2|t|$.  The total number of $\boldsymbol{\kappa}$ taken in the calculation is $N_\kappa=320$; thereby, the Brillouin zone has $\pi/40$ meshes. 

In contrast to the single-particle spectral function, the momentum transfer in two-particle correlation functions such as the dynamical spin correlation function is restricted to discrete momenta defined by the $N$-site lattice even if we introduce the twisted BC.  In order to evaluate various two-particle correlation functions under the twisted BC, we average the correlation functions over the $N_\kappa$ points of $\boldsymbol{\kappa}$.  This procedure is known to reduce finite-size effects.~\cite{Poilblanc}

The dynamical spin correlation function reads
\begin{equation}
S(\mathbf{q},\omega)=\frac{1}{N_\kappa}\sum_{\boldsymbol{\kappa}}\sum_m \left|\left< \Psi^{\boldsymbol{\kappa}}_m \left| S^z_\mathbf{q} \right|\Psi^{\boldsymbol{\kappa}}_0\right>\right|^2 \delta (\omega- E^{\boldsymbol{\kappa}}_m + E^{\boldsymbol{\kappa}}_0)\;,
\label{Sqw}
\end{equation}
where $S^z_\mathbf{q}$ is the Fourier component of the $z$ component of the spin operator with momentum transfer $\mathbf{q}$.  A standard Lanczos technique with a Lorentzian broadening of $0.02|t|$ is used for each $\boldsymbol{\kappa}$ in (\ref{Sqw}).

The two-magnon Raman scattering spectrum with the B$_{1\mathrm{g}}$ symmetry is given by
\begin{equation}
I_\mathrm{R}(\omega)=\frac{1}{N_\kappa}\sum_{\boldsymbol{\kappa}} \sum_m \left|\left< \Psi^{\boldsymbol{\kappa}}_m \left| R \right|\Psi^{\boldsymbol{\kappa}}_0\right>\right|^2 \delta (\omega- E^{\boldsymbol{\kappa}}_m + E^{\boldsymbol{\kappa}}_0)\;
\label{Magnon}
\end{equation}
with the Raman operator for B$_{1\mathrm{g}}$ mode
\begin{equation}
R=\sum_\mathbf{i}(\mathbf{S}_{\mathbf{i}+\mathbf{x}}\cdot\mathbf{S}_\mathbf{i} - \mathbf{S}_{\mathbf{i}+\mathbf{y}}\cdot\mathbf{S}_\mathbf{i})\;.
\label{Raman}
\end{equation}
A standard Lanczos technique with a Lorentzian broadening of $0.1|t|$ is used for each $\boldsymbol{\kappa}$ in (\ref{Raman}).

The real part of the optical conductivity under the electric field applied along the $x$ direction is given by
\begin{equation}
\sigma(\omega)=\sigma_\mathrm{sing}(\omega)+\sigma_\mathrm{reg}(\omega)\;,
\label{SigmaOmega}
\end{equation}
where the regular part $\sigma_\mathrm{reg}(\omega)$ is 
\begin{equation}
\sigma_\mathrm{reg}(\omega)=\frac{1}{N_\kappa}\sum_{\boldsymbol{\kappa}} \frac{\pi}{N\omega} \sum_m \left|\left< \Psi^{\boldsymbol{\kappa}}_m \left| j_x \right|\Psi^{\boldsymbol{\kappa}}_0\right>\right|^2 \delta (\omega- E^{\boldsymbol{\kappa}}_m + E^{\boldsymbol{\kappa}}_0)\;
\label{Regular}
\end{equation}
with the $x$ component of the current operator
\begin{eqnarray}
j_x&=&-\mathrm{i}
 t \sum_{\mathbf{i},\sigma}
    \left( \tilde{c}_{\mathbf{i}+\mathbf{x},\sigma }^\dagger \tilde{c}_{\mathbf{i},\sigma}-\tilde{c}_{\mathbf{i}-\mathbf{x},\sigma }^\dagger \tilde{c}_{\mathbf{i},\sigma} \right) \nonumber \\
&&  - \mathrm{i} t' \sum_{\mathbf{i},\boldsymbol{\delta}',\sigma}
   \left( \tilde{c}_{\mathbf{i}+\boldsymbol{\delta}',\sigma }^\dagger \tilde{c}_{\mathbf{i},\sigma }-\tilde{c}_{\mathbf{i}-\boldsymbol{\delta}',\sigma}^\dagger \tilde{c}_{\mathbf{i},\sigma } \right) \nonumber \\
&&  - \mathrm{i} 2t'' \sum_{\mathbf{i},\sigma}
   \left( \tilde{c}_{\mathbf{i}+2\mathbf{x},\sigma }^\dagger \tilde{c}_{\mathbf{i},\sigma }-\tilde{c}_{\mathbf{i}-2\mathbf{x},\sigma }^\dagger \tilde{c}_{\mathbf{i},\sigma } \right)\;.
\label{jx}
\end{eqnarray}
The singular part $\sigma_\mathrm{sing}(\omega)$ in (\ref{SigmaOmega}) is related to the charge stiffness $D$, which is sometimes called the Drude weight, through
\begin{equation}
\sigma_\mathrm{sing}(\omega)=2\pi D \delta(\omega)\;.
\label{Singular}
\end{equation}
The $D$ satisfies a sum rule
\begin{eqnarray}
K&=&-\frac{1}{2N}\frac{1}{N_\kappa}\sum_{\boldsymbol{\kappa}}\left< \Psi^{\boldsymbol{\kappa}}_0 \left| \tau_{xx} \right|\Psi^{\boldsymbol{\kappa}}_0\right> \nonumber \\
&=& D+\frac{1}{\pi}\int_0^\infty\sigma_\mathrm{reg}(\omega)\mathrm{d}\omega\;,
\label{D}
\end{eqnarray}
where the stress-tensor operator $\tau_{xx}$ is given by
\begin{eqnarray}
\tau_{xx}&=& -t \sum_{\mathbf{i},\sigma}
    \left( \tilde{c}_{\mathbf{i}+\mathbf{x},\sigma }^\dagger \tilde{c}_{\mathbf{i},\sigma}+\tilde{c}_{\mathbf{i}-\mathbf{x},\sigma }^\dagger \tilde{c}_{\mathbf{i},\sigma} \right) \nonumber \\
&&  - t' \sum_{\mathbf{i},\boldsymbol{\delta}',\sigma}
   \left( \tilde{c}_{\mathbf{i}+\boldsymbol{\delta}',\sigma }^\dagger \tilde{c}_{\mathbf{i},\sigma }+\tilde{c}_{\mathbf{i}-\boldsymbol{\delta}',\sigma}^\dagger \tilde{c}_{\mathbf{i},\sigma } \right) \nonumber \\
&&  -2t'' \sum_{\mathbf{i},\sigma}
   \left( \tilde{c}_{\mathbf{i}+2\mathbf{x},\sigma }^\dagger \tilde{c}_{\mathbf{i},\sigma }+\tilde{c}_{\mathbf{i}-2\mathbf{x},\sigma }^\dagger \tilde{c}_{\mathbf{i},\sigma } \right)\;.
\label{tauxx}
\end{eqnarray}
A standard Lanczos technique with a Lorentzian broadening of $0.1|t|$ is used for each $\boldsymbol{\kappa}$ in calculating (\ref{Regular}).  The same broadening is used for the singular part (\ref{Singular}).

We also calculate static correlation functions as a function of distance $r$.  The spin correlation with staggered phase factors, which is a measure of the strength of AF correlation, is defined as 
\begin{equation}
C_\mathrm{spin}(r)=\frac{1}{N_\kappa}\sum_{\boldsymbol{\kappa}} \frac{P_r}{NN_r} \sum_{\mathbf{i},\boldsymbol{\rho}} \left< \Psi^{\boldsymbol{\kappa}}_0 \left| \mathbf{S}_{\mathbf{i}+\boldsymbol{\rho}}\cdot\mathbf{S}_\mathbf{i} \right|\Psi^{\boldsymbol{\kappa}}_0\right>\;,
\label{Cspin}
\end{equation}
where the summation of $\boldsymbol{\rho}$ is taken to be bonds satisfing $\left|\boldsymbol{\rho}\right|=r$, and $N_r$ is the number of the bonds.  The factor $P_r$ is 1 when the two sites are in the same sublattice, and $-$1 otherwise.    The charge correlation for doped carriers is
\begin{eqnarray}
C_\mathrm{charge}(r)&=&\frac{1}{N_\kappa}\sum_{\boldsymbol{\kappa}} \frac{1}{NN_r} \nonumber \\
&\times& \sum_{\mathbf{i},\boldsymbol{\rho}} \left< \Psi^{\boldsymbol{\kappa}}_0 \left| (1-n_{\mathbf{i}+\boldsymbol{\rho}})(1-n_\mathbf{i}) \right|\Psi^{\boldsymbol{\kappa}}_0\right>\;,
\label{Ccharge}
\end{eqnarray}
where the number operator $n_\mathbf{i}$ is given by $n_\mathbf{i}=\sum_\sigma c^\dagger_{\mathbf{i},\sigma}c_{\mathbf{i},\sigma}$.  In terms of superconductivity in the $t$-$t'$-$t''$-$J$ model, the $d$-wave pairing correlation can be calculated, which is defined as
\begin{equation}
C_\mathrm{pair}(r)=\frac{1}{N_\kappa}\sum_{\boldsymbol{\kappa}} \frac{1}{NN_r} \sum_{\mathbf{i},\boldsymbol{\rho}} \left< \Psi^{\boldsymbol{\kappa}}_0 \left| \Delta^\dagger_{\mathbf{i}+\boldsymbol{\rho}} \Delta_\mathbf{i} \right|\Psi^{\boldsymbol{\kappa}}_0\right>\;,
\label{Cpair}
\end{equation}
where $\Delta_\mathbf{i}$ is the $d_{x^2-y^2}$-wave singlet operator 
\begin{eqnarray}
\Delta_\mathbf{i}&=&\frac{1}{\sqrt{2}}\sum_\sigma (-1)^\sigma (\tilde{c}_{\mathbf{i},\sigma}\tilde{c}_{\mathbf{i}+\mathbf{x},-\sigma}+\tilde{c}_{\mathbf{i},\sigma}\tilde{c}_{\mathbf{i}-\mathbf{x},-\sigma} \nonumber \\
& &  -\tilde{c}_{\mathbf{i},\sigma}\tilde{c}_{\mathbf{i}+\mathbf{y},-\sigma}-\tilde{c}_{\mathbf{i},\sigma}\tilde{c}_{\mathbf{i}-\mathbf{y},-\sigma})\;.
\label{d}
\end{eqnarray}

\section{Magnetic properties}
\label{MagneticProperties}

In this section we show calculated results of doping dependences of the spin correlation functions and two-magnon Raman scattering for the $N=20$ square lattice of the $t$-$t'$-$t''$-$J$ model.

\begin{figure}
\begin{center}
\includegraphics[width=8.5cm]{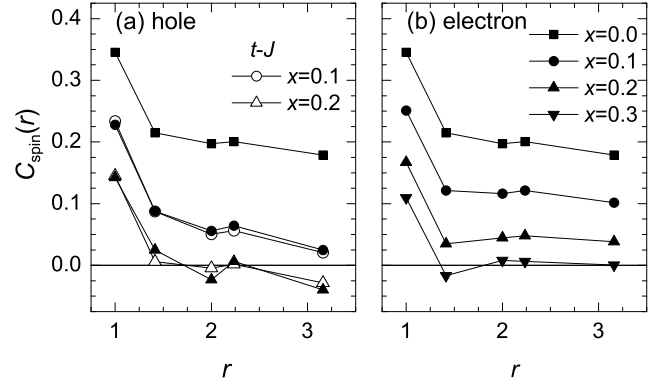}
\caption{\label{CspinFIG}
Staggered spin correlation $C_\mathrm{spin}(r)$ as a function of the two-spin distance $r$ for an $N=20$ $t$-$t'$-$t''$-$J$ lattice, obtained by averaging over twisted BC.  (a) Hole doping ($t=1$, $t'=-0.25$, $t''=0.12$, and $J=0.4$) and  (b) electron doping ($t=-1$, $t'=0.25$, $t''=-0.12$, and $J=0.4$).  Solid squares, circles, upper triangles, and lower triangles are $x=0$ (Heisenberg model), $x=0.1$, $x=0.2$, and $x=0.3$, respectively.  For comparison the $t$-$J$ results are plotted with open symbols in (a).}
\end{center}
\end{figure}

Figure~\ref{CspinFIG} shows the staggered spin correlation $C_\mathrm{spin}(r)$ in (\ref{Cspin}) for different carrier concentration $x$, where $x=N_\mathrm{c}/N$, $N_\mathrm{c}$ being the number of carriers in the $N$-site lattice.  For the Heisenberg model ($x=0.0$), the correlation is almost constant at large distance, indicating the presence of AF order.  In hole doping, the AF correlation is suppressed and rapidly decays at large distance with increasing $x$.  On the other hand, the AF correlation in electron doping is similar to that of the Heisenberg model, although the magnitude decreases with doping. This indicates the presence of AF order even in the electron doped systems up to $x=0.2$.  At $x=0.3$ the AF correlation almost disappears at large distance.  We note that, although detailed studies on doping dependence of the magnetic correlation length for electron-doped NCCO have been reported,~\cite{Matsuda,Mang} the present results cannot be compared with them because of small system size.  For a $t$-$t'$-$t''$-$U$ model, where $U$ represents the on-site Coulomb interaction, it has been reported~\cite{Kyung2} that the temperature-dependent correlation length agrees with the experiments.

The difference of AF correlation between hole and electron dopings in the $t$-$t'$(-$t''$)-$J$ model has already been pointed out under periodic BC.~\cite{Tohyama1,Gooding,Tohyama2}  The present results obtained by averaging over twisted BC confirm that such a difference in AF correlation is independent of BC and thus intrinsic to the model.  The origin of the difference comes from the sign difference of $t'$ and $t''$.~\cite{Tohyama1,Tohyama2}  Let us consider the Hilbert-space bases with N\'eel-type spin configuration in the spin background.   The $t'$ and $t''$ do not change spin configuration of these bases because of the same sublattice hoppings.  This means that the self-energies of the bases are dependent on the values of $t'$ and $t''$.  We can find~\cite{Tohyama1,Tohyama2} that the energies become lower when $t'>0$ corresponding to electron doping.  This stabilizes the N\'eel-type spin configuration; thereby, AF correlation remains strong in electron doping.  For hole doping, on the other hand, the self-energies of the bases are increased and thus become comparable with other bases with different spin configurations.  Such a mixture of various spin configurations gives rise to a similar effect caused by nearest-neighbor hopping of carriers.  In fact, we find from Fig.~\ref{CspinFIG}(a) that the spin correlation is similar to that of the $t$-$J$ model. We note that the similarity between the $t$-$t'$-$t''$-$J$ and $t$-$J$ model has not been observed in a cluster under the periodic BC, where the spin correlation decreases with increasing the magnitude of $t'$ and $t''$.

\begin{figure}
\begin{center}
\includegraphics[width=8.5cm]{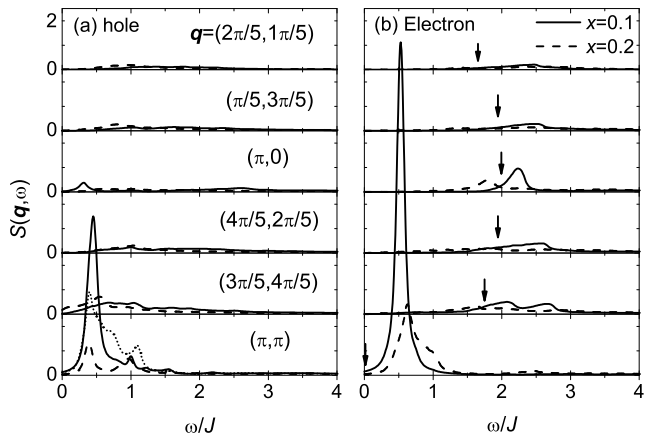}
\caption{\label{SqwFIG}
Dynamical spin correlation function $S(\mathbf{q},\omega)$ for an $N=20$ $t$-$t'$-$t''$-$J$ lattice, obtained by averaging over twisted BC.  (a) Hole doping ($t=1$, $t'=-0.25$, $t''=0.12$, and $J=0.4$) and  (b) electron doping ($t=-1$, $t'=0.25$, $t''=-0.12$, and $J=0.4$).  Solid and dashed lines represent the data for $x=0.1$ and $x=0.2$, respectively.  The dotted line at $\mathbf{q}=(\pi,\pi)$ in (a) represents the data for the $t$-$J$ model at $x=0.1$.  The momenta defined in the lattice are shown in (a).  In (b) the edge of the spin-wave excitations in the Heisenberg model ($x=0$) obtained by the linear-spin-wave theory is indicated by the downward arrow for each momentum.}
\end{center}
\end{figure}

The dynamical spin correlation functions (\ref{Sqw}) in the $N=20$ lattice are shown in Fig.~\ref{SqwFIG}.  For $x=0.1$ of electron doping, the excitation at $\mathbf{q}=(\pi,\pi)$ exhibits the minimum energy among the momenta defined in the lattice, and has the largest weight.  Since the staggered spin correlation $C_\mathrm{spin}(r)$ shown in Fig.~\ref{CspinFIG} indicates the presence of long-range AF order, the finite-excitation energy at $\mathbf{q}=(\pi,\pi)$ can be due to the finite-size effect that inevitably causes a discrete energy separation between the ground state and excited states.  Away from $(\pi,\pi)$, the spectral weights are distributed at the higher-energy region, whose scale is comparable with the spin-wave excitation of the Heisenberg model whose lower-bound edges are denoted by the downward arrows.  This again confirms that the ground state is the AF ordered state even in the presence of mobile carriers.  With further doping of electrons ($x=0.2$), the $(\pi,\pi)$ spectrum loses its weight and the high-energy weights at other momenta shift to the lower-energy side, as expected from the reduction of AF correlation.  This doping dependence is qualitatively consistent with recent inelastic neutron scattering measurements for an electron-doped material, Pr$_{2-x}$Ce$_x$CuO$_4$.~\cite{Fujita1}

At $x=0.1$ of hole doping, the lowest-energy excitations are not at $\mathbf{q}=(\pi,\pi)$ but at $(\pi,0)$ as shown in Fig.~\ref{SqwFIG}(a), although the spectral weight is the highest at $(\pi,\pi)$.  For $x=0.2$, the $(\pi,\pi)$ weight decreases and the weight at $(3\pi/5,4\pi/5)$ shifts to lower energy.  These behaviors indicate a tendency toward incommensurate spin correlations reported in hole-doped materials such as LSCO (Ref.~8) and YBa$_2$Cu$_3$O$_{7-\delta}$.~\cite{Reznik}  At $(\pi,\pi)$ for $x=0.1$, we find, from the comparison between the solid and dotted lines, that the introduction of $t'$ and $t''$ shifts weights to lower-energy side.  Since the equal-time spin correlations are almost the same, as seen in Fig.~\ref{CspinFIG}, we can say that the effect of $t'$ and $t''$ on magnetic properties is the shift of AF spin fluctuation toward lower frequencies.   

\begin{figure}
\begin{center}
\includegraphics[width=8.5cm]{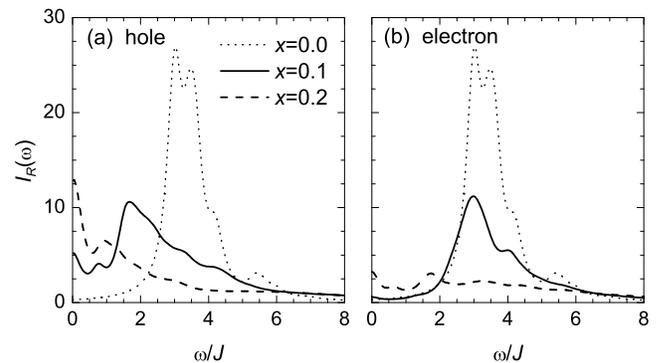}
\caption{\label{RamanFIG}
Two-magnon Raman spectrum $I_R(\omega)$ with B$_{1\mathrm{g}}$ geometry for an $N=20$ $t$-$t'$-$t''$-$J$ lattice, obtained by averaging over twisted BC.  (a) Hole doping ($t=1$, $t'=-0.25$, $t''=0.12$, and $J=0.4$) and  (b) electron doping ($t=-1$, $t'=0.25$, $t''=-0.12$, and $J=0.4$).  Solid and dashed lines represent the data for $x=0.1$ and $x=0.2$, respectively.  The dotted lines  represent the spectrum in the Heisenberg model ($x=0$) obtained by averaging over the spectra for $N=$16, 18, 20, and 26 lattices.}
\end{center}
\end{figure}

Two-magnon Raman scattering also shows such contrasting magnetic behaviors between hole- and electron-doped systems.  Figure~\ref{RamanFIG} exhibits the B$_{1\mathrm{g}}$ two-magnon Raman spectrum $I_R(\omega)$ in (\ref{Raman}) for the $N=20$ lattice.   The dotted lines in the figure represent the B$_{1\mathrm{g}}$ two-magnon Raman spectrum of the Heisenberg model ($x=0$), which is obtained  by averaging over the spectra for $N=$16, 18, 20, and 26 lattices in order to reduce finite-size effects.  The main peak due to two magnons appears at around $\omega\sim3J$.~\cite{Sandvik}  For $x=0.1$ of electron doping, the two-magnon peak position remains unchanged with $\omega\sim3J$ as expected from the presence of AF order, though its weight decreases.  In contrast, the two-magnon peak shifts to the lower-energy side in hole doping from $\omega\sim3J$ to $2J$ at $x=0.1$.  At $x=0.2$, there is no pronounced magnon peaks in both the hole and electron dopings.  Such a contrasting behavior about the position of the two-magnon peak is consistent with recent experimental data, where in hole doping the peak position shifts to the lower-energy side~\cite{Sugai1} but not in electron doping.~\cite{Sugai2}  We note that the line shapes of the calculated spectra cannot be directly compared with those of experimental ones because of excluding electronic Raman contributions.

\section{Single-particle spectral function}
\label{SpectralFunction}

In this section, we present doping dependence of the single-particle spectral function $A(\mathbf{k},\omega)$ for both hole- and electron-doped $t$-$t'$-$t''$-$J$ models, and discuss the similarity and difference between the calculated results and ARPES data.

\subsection{Half filling}
\label{HalfFilling}

\begin{figure}
\begin{center}
\includegraphics[width=8.cm]{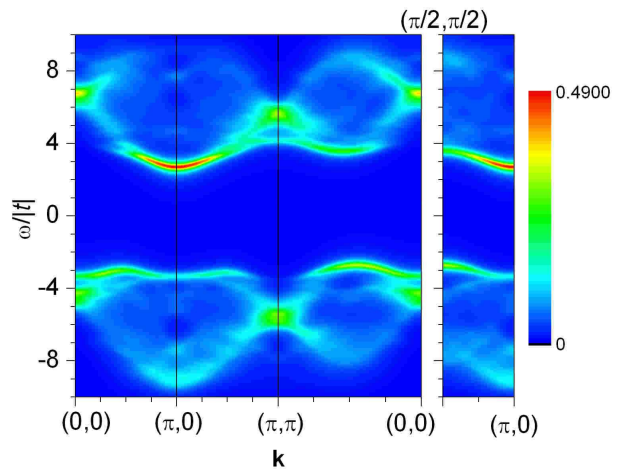}
\caption{\label{Akwx=0}
(Color online) Weight map of the spectral function for an $N=20$ $t$-$t'$-$t''$-$J$ model at half filling along the high-symmetry lines. $|t|=1$, $t'/t=-0.25$, $t''/t=0.12$, and $J/|t|=0.4$.   Twisted BC are imposed on the lattice in calculating the final states.  For each BC a Lorentzian broadening of $0.2|t|$ is used.  The scale of the weight is shown in the bar at the right side of the panels.  The on-site Coulomb interaction that determines the Mott-gap magnitude is set to be $U/|t|=4|t|/J=10$.  The chemical potential is located at the zero energy.}
\end{center}
\end{figure}

Let us start with the spectral function at half filling ($x=0$).  Figure~\ref{Akwx=0} shows the weight map along the high-symmetry lines in the first Brillouin zone obtained by introducing the twist for the $N=20$ $t$-$t'$-$t''$-$J$ model as mentioned in Sec.~\ref{Model}.  To obtain continuous weights, we perform a smoothing procedure for each symmetry line that has $\pi/40$ meshes.~\cite{Xiang}  The on-site Coulomb interaction that determines Mott-gap magnitude is set to be $U/|t|=4|t|/J=10$.

The top of the lower Hubbard band is located at $\mathbf{k}=(\pi/2,\pi/2)$.  Quasiparticle energies at around $(\pi,0)$ are lower than that of $(\pi/2,\pi/2)$.  The spectral weights at around $(\pi,0)$ are suppressed in contrast to the case of the $t$-$J$ model, where the quasiparticle energies at both $\mathbf{k}=(\pi,0)$ and $(\pi/2,\pi/2)$ are almost degenerate and their weights are similar.~\cite{Tohyama4}  On the other hand, the quasiparticle at $(\pi,0)$ in the upper Hubbard band is located at the bottom of the band.  Therefore, the charge excitation with minimum energy is from $\mathbf{k}=(\pi/2,\pi/2)$ in the lower Hubbard band to $(\pi,0)$ in the upper Hubbard band.  In other words, the Mott gap in the two-dimensional insulating cuprates is an indirect gap, as previously pointed out by the exact diagonalization study of a Hubbard model with $t'$ and $t''$.~\cite{Tsutsui2}  From such an indirect nature, we can expect that doped holes predominantly enter into the $(\pi/2,\pi/2)$ region in heavy underdoping, while the electrons enter into the $(\pi,0)$ region.

We also see in Fig.~\ref{Akwx=0} that the spectral weights at around the bottom of the upper Hubbard band are the largest among those at other regions.  Since spectral weight at half filling is roughly proportional to the product of the weight of the N\'eel-type configuration in the Heisenberg ground state by that in the single-carrier final state, the large weights at around $(\pi,0)$ imply that the N\'eel-type configuration is dominant in the final states near $(\pi,0)$, and thus AF correlation is strong in the low-energy sectors of the single electron-doped system.  Such a momentum-dependent feature comes from the presence of $t'$ and $t''$.~\cite{Tohyama2}

\begin{figure}
\begin{center}
\includegraphics[width=8.cm]{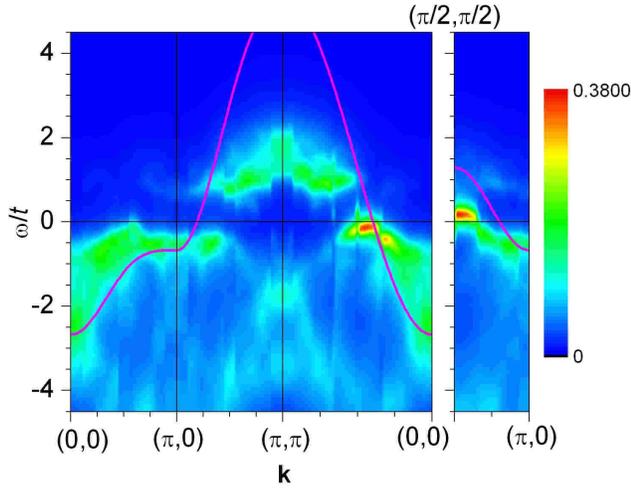}
\caption{\label{Akwx=0.1hole}
(Color online) Weight map of the spectral function for an $N=20$ $t$-$t'$-$t''$-$J$ model at the hole concentration $x=1-18/20=0.1$.  $t=1$, $t'=-0.25$, $t''=0.12$, and $J=0.4$.   Twisted BC are imposed on the lattice.  For each BC a Lorentzian broadening of $0.2t$ is used.  The scale of the weight is shown in the bar at the right side of the panels.  The red curves represent a noninteracting tight-binding band with the same hopping amplitudes.}
\end{center}
\end{figure}

\subsection{Hole doping}
\label{HoleDoping}

Figure~\ref{Akwx=0.1hole} shows the weight map of the spectral function for a two-hole doped system ($x=0.1$) along the high-symmetry lines in the first Brillouin zone.   The red lines in the figure represent a noninteracting tight-binding band with the same hopping amplitudes as $t=1$, $t'/t=-0.25$, and $t''/t=0.12$, which is available for a guide of the band renormalization of the $t$-$t'$-$t''$-$J$ model.  From the weight map, we find that large spectral weights appear at around $\mathbf{k}=(\pi/2,\pi/2)$ just below and above the Fermi level.  We also find that along the $(0,0)$-$(\pi,\pi)$ direction the dispersion exhibits a slight downturn toward $(0,0)$ at $(\pi/2,\pi/2)$.  On the other hand, at around $(\pi,0)$ the spectra are located below the Fermi level with small weight and flat dispersion.  These behaviors are consistent with the picture that doped holes predominantly occupy the $(\pi/2,\pi/2)$ region in underdoped system, which is expected from the dispersion at half filling as discussed above.

\begin{figure}
\begin{center}
\includegraphics[width=8.5cm]{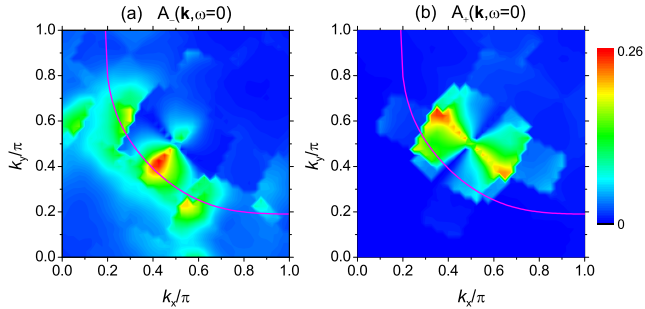}
\caption{\label{FSx=0.1hole}
(Color online) Contour plot of $A(\mathbf{k},\omega=0)$ for an $N=20$ $t$-$t'$-$t''$-$J$ model at the hole concentration $x=1-18/20=0.1$.  $t=1$, $t'=-0.25$, $t''=0.12$, and $J=0.4$.  (a) Electron-removal spectrum $A_-(\mathbf{k},\omega=0)$ and (b) electron-addition spectrum $A_+(\mathbf{k},\omega=0)$.  The scale of the weight is shown in the bar at the right side of the panels.  The red curves represents a noninteracting tight-binding Fermi surface with the same hopping amplitudes.}
\end{center}
\end{figure}

The most interesting feature in Fig.~\ref{Akwx=0.1hole} is a gapped behavior near the Fermi level along the $(\pi,0)$-$(\pi,\pi)$ direction.  This seems to correspond to the pseudogap observed in ARPES experiments for hole-doped high-$T_c$ cuprates.~\cite{Damascelli}  In order to see such a gap feature in more detail, we show in Fig.~\ref{FSx=0.1hole} the intensity map at the Fermi level ($\omega=0$) for both the electron-removal ($A_-$) and electron-addition ($A_+$) spectra.  Note that in finite-size lattices $A_-(\mathbf{k},\omega=0)$ is not equal to $A_+(\mathbf{k},\omega=0)$, in contrast with the case of the thermodynamic limit, and rather $A_-(\mathbf{k},\omega=0)$  in this case is comparable with ARPES intensity near the Fermi level.  In Fig.~\ref{FSx=0.1hole}(a), spectral weights are large near the nodal $(0,0)$-$(\pi,\pi)$ direction, and the weights decrease away from the nodal region along the original noninteracting Fermi surface.  This is similar to the so-called {\it Fermi arc} observed in the normal state of underdoped high-$T_c$ materials.~\cite{Yoshida,Ronning,Damascelli}  We do not find pocket-like features in the electron-removal side at around $(\pi/2,\pi/2)$, consistent with the experiments.

\begin{figure}
\begin{center}
\includegraphics[width=8.cm]{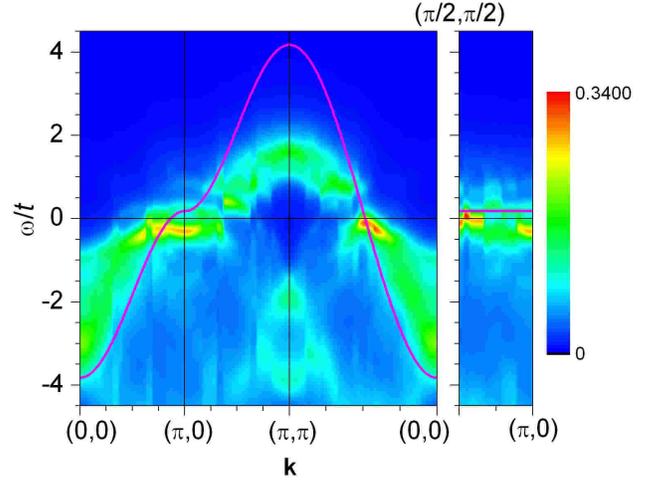}
\caption{\label{AkwtJx=0.1hole}
(Color
 online) Same as Fig.~\ref{Akwx=0.1hole} but for a $t$-$J$ model.}
\end{center}
\end{figure}

We also note that the ridge of the spectra in Fig.~\ref{FSx=0.1hole}(a) reaches the momenta of $(0.6\pi,0)$ and $(0,0.6\pi)$.  Such a feature has experimentally observed in Na-doped Ca$_2$CuO$_2$Cl$_2$,~\cite{Ronning} and theoretically demonstrated by using the method of equations of motion for the $t$-$t'$-$J$ model.~\cite{Prelovsek}  In the present calculations, the presence of the weight at around $(0.6\pi,0)$ and $(0,0.6\pi)$ is interpreted as the effect of $t'$ and $t''$: The quasiparticle energy at around $(\pi/2,0)$ is insensitive to $t'$ and $t''$ in contrast with that at $(\pi,0)$, as is expected from the tight-binding form.  Thereby, the quasiparticle is kept close to the Fermi level at the $(\pi/2,0)$ region.  

In the electron-addition side [Fig.~\ref{FSx=0.1hole}(b)], spectral weights predominantly spread along the $(0,\pi)$-$(\pi,0)$ direction.  In order to detect these weights, we need angle-resolved inverse photoemission with high resolution.  

The gap along the ($\pi$,0)-($\pi$,$\pi$) direction is sensitive to the magnitude of $t'$ and $t''$.  To see this, we show in Fig.~\ref{AkwtJx=0.1hole} the spectral function for the $t$-$J$ model with $x=0.1$.  We find a less clear gap observed along the ($\pi$,0)-($\pi$,$\pi$) direction, although the dispersion along the $(0,0)$-$(\pi,\pi)$ direction is similar to that of the $t$-$t'$-$t''$-$J$ model.  Therefore, we can say that the long-range hoppings $t'$ and $t''$ are responsible for the formation of the gap and thus for a Fermi-arc behavior observed in ARPES experiments of hole-doped cuprates.~\cite{Yoshida,Ronning,Damascelli}  Recent ARPES experiments~\cite{Tanaka} have shown that the flat band at around $\mathbf{k}=(\pi,0)$ is deeper in energy for Bi$_{2-z}$Pb$_z$Sr$_2$Ca$_{1-x}$(Pr,Er)$_x$Cu$_2$O$_{8+\delta}$ (BSCCO) than for LSCO.  Since $t'$ and $t''$ for BSCCO are known to be larger than that for LSCO,~\cite{Tohyama4} the experimental data are consistent with the present picture that $t'$ and $t''$ predominantly control the pseudogap magnitude.

\begin{figure}
\begin{center}
\includegraphics[width=7.5cm]{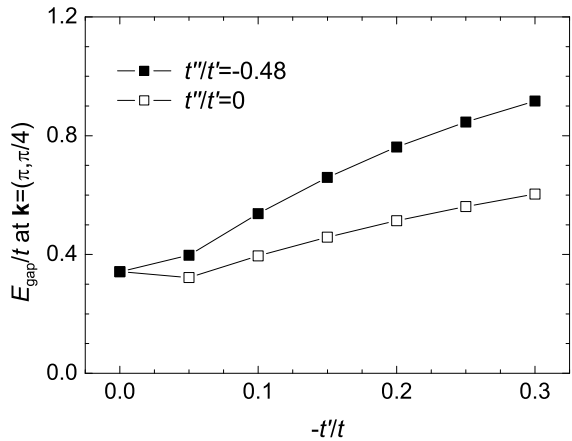}
\caption{\label{Gap_pi_pi/4}
Dependence of the gap energy $E_\mathrm{gap}$ on $t'$ for an $N=20$ $t$-$t'$-$t''$-$J$ model at the hole concentration $x=0.1$.  $t=1$ and $J=0.4$.  $E_\mathrm{gap}$ is the minimum-energy difference between the electron-removal and electron-addition states at $\mathbf{k}=(\pi,\pi/4)$.  The ratio of $t''/t$ is kept at $-0.12/0.25=-0.48$ (solid squares) and $0$ (open squares).}
\end{center}
\end{figure}

In order to examine this more quantitatively, we plot in Fig.~\ref{Gap_pi_pi/4} the $t'$ dependence of the gap energy $E_\mathrm{gap}$ for two cases of $t''/t'=-0.12/0.25=-0.48$ and $t''/t'=0$.  The $E_\mathrm{gap}$ is defined as the minimum-energy difference between the electron-removal and electron-addition states at $\mathbf{k}=(\pi,\pi/4)$.  At $-t'/t=0$, the gap is finite but may come partly from the finite-size effect.  With increasing $-t'/t$, $E_\mathrm{gap}$ increases for both the cases of $t''/t'=0$ and $-0.48$.  Compairing them, we also find that $t''$ significantly contributes for the gap magnitude.  Since $t'$ and $t''$ not only reduce quasiparticle energy at around $(\pi,0)$ but also enhance low-energy AF fluctuation in the spin background, the magnitude of the gap seems to be related to the strength of AF fluctuation.  In fact, we find that the gap is also dependent on the value of $J$: The gap at $-t'/t=0.25$ changes from $0.85t$ for $J/t=0.4$ to $0.51t$ for $J/t=0.2$. 

\begin{figure}
\begin{center}
\includegraphics[width=8.cm]{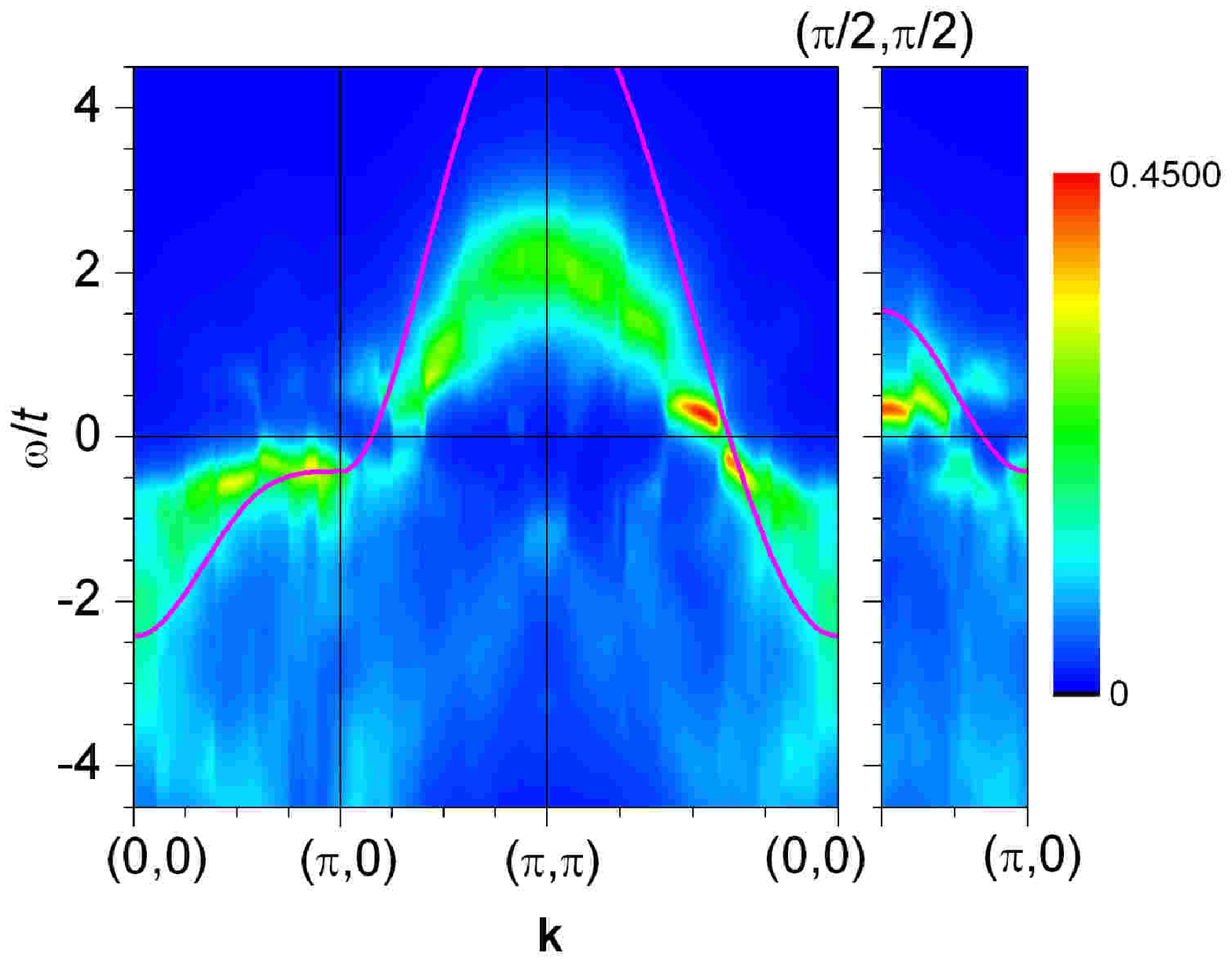}
\caption{\label{Akwx=0.2hole}
(Color online) Same as Fig.~\ref{Akwx=0.1hole} but $x=1-16/20=0.2$.}
\end{center}
\end{figure}

From Fig.~\ref{Akwx=0.1hole} we see that gap opens not only along $(\pi,0)$-$(\pi,\pi)$ but also along the $(0,0)$-$(\pi,\pi)$ direction.  In the latter, the gap appears in the electron-addition side ($\omega>0$), and its magnitude is smaller than that at $(\pi,\pi/4)$.~\cite{Prelovsek}  The gap value is also  found to be dependent on $t'$ if Fig.~\ref{Akwx=0.1hole} is compared with Fig.~\ref{AkwtJx=0.1hole}.  In order to detect this gap, we need angle-resolved inverse photoemission experiments.

At $x=0.2$ (four holes in the $N=20$ lattice), the gap becomes less clear than that at $x=0.1$, as shown in Fig.~\ref{Akwx=0.2hole}.  Along the $(0,0)$-$(\pi,\pi)$ direction, the gap feature almost disappears, and a downturn  of the dispersion at $(\pi/2,\pi/2)$, seen at $x=0.1$, vanishes completely.  Furthermore, a spectral distribution near the Fermi level along the $(\pi,0)$-$(\pi,\pi)$ direction becomes continuous, though the weight at around $(\pi,\pi/4)$ is still small.  We can say that overall behavior gradually approaches the noninteracting band with increasing hole concentration.

\subsection{Electron doping}
\label{ElectronDoping}

\begin{figure}
\begin{center}
\includegraphics[width=8.cm]{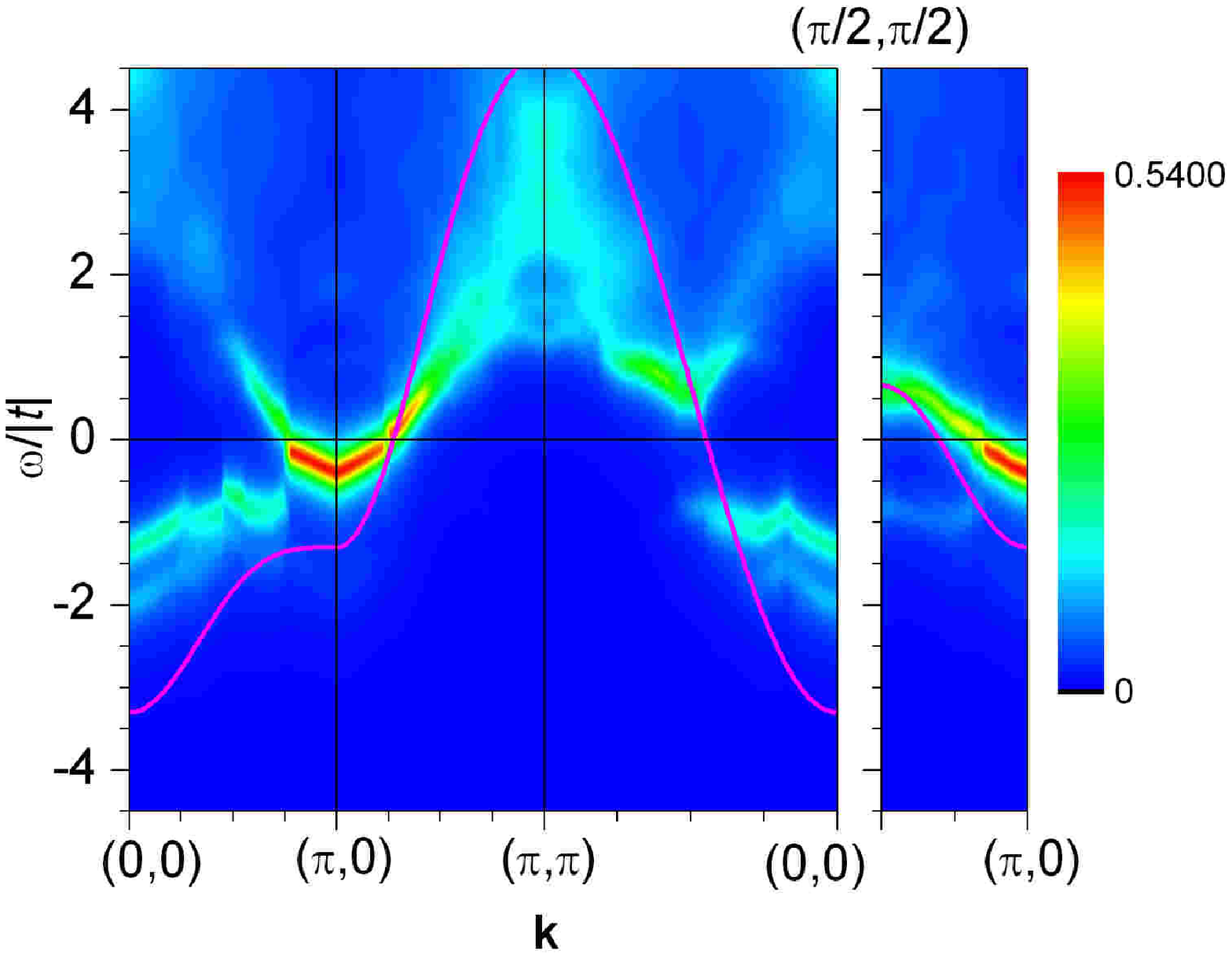}
\caption{\label{Akwx=0.1ele}
(Color online) Weight map of the spectral function for an $N=20$ $t$-$t'$-$t''$-$J$ model at the electron concentration $x=22/20-1=0.1$.  $t=-1$, $t'=0.25$, $t''=-0.12$, and $J=0.4$.  Twisted BC are imposed on the lattice.  For each BC a Lorentzian broadening of $0.2|t|$ is used.  The scale of the weight is shown in the bar at the right side of the panels.  The red curves represent a noninteracting tight-binding band with the same hopping amplitudes.}
\end{center}
\end{figure}

Figure~\ref{Akwx=0.1ele} shows $A(\mathbf{k},\omega)$ for a two-electron doped system ($x=0.1$) of the $N=20$ $t$-$t'$-$t''$-$J$ model.  The spectra are very different from those for hole doping.  At around $\mathbf{k}=(\pi,0)$, an electron pocket is seen as expected from the spectral function at half filling (see Fig.~\ref{Akwx=0}).  Note that, in order to get such a clear pocket, the presence of AF order is necessary.~\cite{Tohyama2}  In fact, strong AF correlation remains in the lattice, as explained in Sec.~\ref{MagneticProperties}.  Along the $(0,0)$-$(\pi,\pi)$ direction, we find a clear gap in the dispersion.

\begin{figure}
\begin{center}
\includegraphics[width=8.5cm]{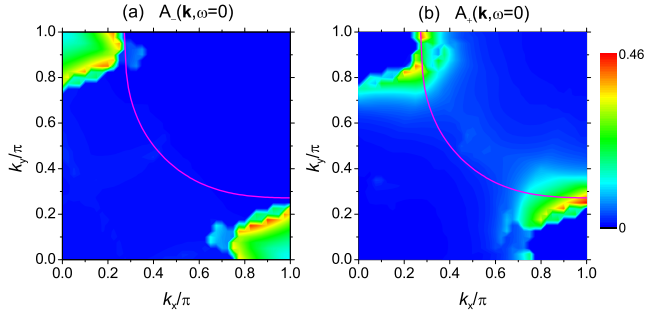}
\caption{\label{FSx=0.1ele}
(Color online) Contour plot of $A(\mathbf{k},\omega=0)$ for an $N=20$ $t$-$t'$-$t''$-$J$ model at the electron concentration $x=22/20-1=0.1$.  $t=-1$, $t'=0.25$, $t''=-0.12$, and $J=0.4$.  (a) Electron-removal spectrum $A_-(\mathbf{k},\omega=0)$ and (b) electron-addition spectrum $A_+(\mathbf{k},\omega=0)$.  The red curves represents a noninteracting tight-binding Fermi surface with the same hopping amplitudes.}
\end{center}
\end{figure}

The Fermi surface map is shown in Fig.~\ref{FSx=0.1ele}, where electron pockets centered at $(\pi,0)$ and $(0,\pi)$ are clearly seen.  The presence of the pockets and the absence of weights along the nodal direction are consistent with ARPES data~\cite{Armitage2} for underdoped NCCO where AF long-rang order persists.  We also note that recent spectral function calculations for a Hubbard model with long-range hoppings display gap behaviors consistent with our results as long as $U$ is large.~\cite{Kusuko,Markiewicz,Kyung1,Kyung2,Senechal,Kusunose}

\begin{figure}
\begin{center}
\includegraphics[width=8.cm]{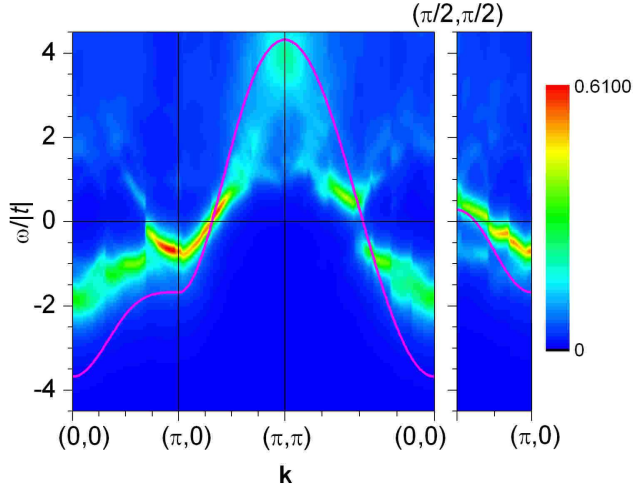}
\caption{\label{Akwx=0.2ele}
(Color online) Same as Fig.~\ref{Akwx=0.1ele} but $x=24/20-1=0.2$.}
\end{center}
\end{figure}

\begin{figure}
\begin{center}
\includegraphics[width=8.cm]{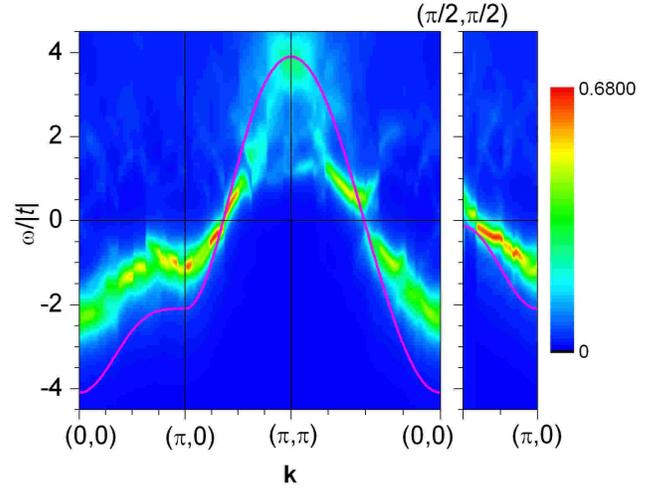}
\caption{\label{Akwx=0.3ele}
(Color online) Same as Fig.~\ref{Akwx=0.1ele} but $x=26/20-1=0.3$.}
\end{center}
\end{figure}

The spectral function for a four-electron doped system ($x=0.2$) is exhibited in Fig.~\ref{Akwx=0.2ele}.  The $(\pi,0)$ electron pocket seen at $x=0.1$ almost disappears, although the spectral intensity at around $(\pi,0)$ is still strong enough to show a remnant of the pocket.  In contrast, the $(\pi/2,\pi/2)$ gap clearly remains but with smaller gap magnitude.  With further doping, the spectra show dispersions similar to a noninteracting system.  Figure~\ref{Akwx=0.3ele} exhibits the case of $x=0.3$, where the gap at around $(\pi/2,\pi/2)$ almost disappears and the dispersion qualitatively follows the noninteracting band.  The velocity at the Fermi level is almost one half of the noninteracting-band velocity, which is independent of the Fermi momentum.

\begin{figure}
\begin{center}
\includegraphics[width=7.5cm]{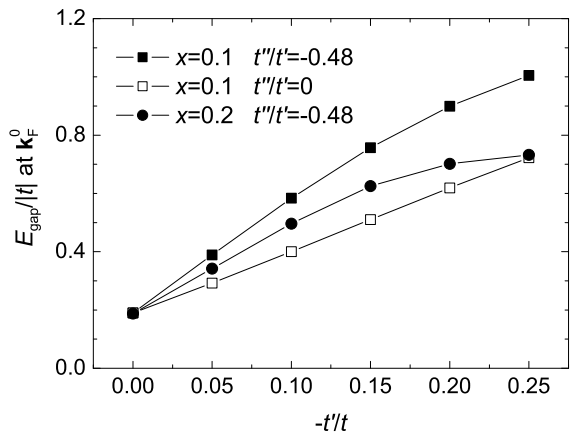}
\caption{\label{Gap_pi/2_pi/2}
Dependence
 of the gap energy $E_\mathrm{gap}$ on $t'$ for an electron-doped $N=20$ $t$-$t'$-$t''$-$J$ model.  $t=-1$ and $J=0.4$.  $E_\mathrm{gap}$ is the the minimum-energy difference between the electron-removal and electron-addition states at the noninteracting Fermi momentum $\mathbf{k}_\mathrm{F}^0$ along the $(0,0)$-$(\pi,\pi)$ direction.  For the electron concentration $x=0.1$, the ratio of $t''/t$ is kept at $-0.12/0.25=-0.48$ (solid squares) and $0$ (open squares).  For $x=0.2$, $t''/t=-0.48$ (solid circles).}
\end{center}
\end{figure}

Let us discuss the origin of the gap along the $(0,0)$-$(\pi,\pi)$ direction seen for $x=0.1$ and $0.2$.  The gap is, of course, the consequence of the presence of $t'$ and $t''$, because there is no gap at the Fermi level along the nodal direction in the $t$-$J$ model (see Fig.~\ref{AkwtJx=0.1hole}).  In Fig.~\ref{Gap_pi/2_pi/2}, we show the $t'$ dependence of the gap energy $E_\mathrm{gap}$ defined at the noninteracting Fermi momentum $\mathbf{k}^0_\mathrm{F}$ along the nodal direction.  Since the $t$-$J$ model has no gap, the gap value of $0.2|t|$ at $-t'/t=0$ is due to the finite-size effect.  As is the case of hole doping (see Fig.~\ref{Gap_pi_pi/4}), we find that $E_\mathrm{gap}$ increases with increasing $-t'/t$ for both $x=0.1$ and $0.2$.  From the comparison between the cases of $t''/t'=0$ and $-0.48$ for $x=0.1$, we also find that $t''$ significantly contributes to the formation of the gap.  Since $t'$ and $t''$ enhance AF correlation as discussed in Sec.~\ref{MagneticProperties}, the magnitude of the gap in electron doping seems to be related to the strength of AF correlation.  Reflecting the sensitivity to the AF correlation, when we increase $J$ from 0.4 to 0.6 keeping $t''/t'=-0.48$ for $x=0.1$, $E_\mathrm{gap}$ increases from $1.01|t|$ to $1.33|t|$.   

Even for $x=0.2$, the gap increases with increasing $-t'/t$.  This implies that AF correlation is still strong enough to produce the gap.  However, comparing with the experimental fact that at this concentration the $d$-wave superconductivity emerges at low temperature, the gap obtained in the present calculation may indicate an overestimate of the AF correlation.  Thus, we may need to clarify the mechanism of the gap closing which makes the system a $d$-wave superconductor.  We will discuss this in the following section.

\section{Charge dynamics and pairing properties}
\label{ChargeDynamics}

In this section, we first discuss the difference of the optical conductivity between the hole- and electron-doped $t$-$t'$-$t''$-$J$ models.  Next, we show the $d$-wave pairing correlations together with charge correlations in the models.

\subsection{Optical conductivity}
\label{OpticalConductivity}

\begin{figure}
\begin{center}
\includegraphics[width=8.5cm]{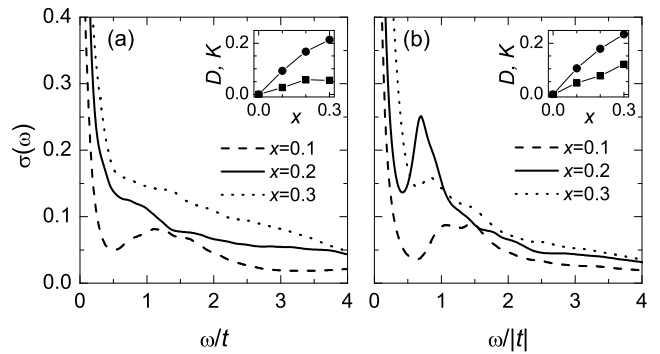}
\caption{\label{OptFig}
Optical conductivity $\sigma(\omega)$ for an $N=20$ $t$-$t'$-$t''$-$J$ model.  (a) Hole doping ($t=1$, $t'=-0.25$, $t''=0.12$, and $J=0.4$) and (b) electron doping ($t=-1$, $t'=0.25$, $t''=-0.12$, and $J=0.4$).  Dashed, solid, and dotted lines represent the carrier concentration of $x=0.1$, $0.2$, and $0.3$, respectively.  Delta functions are broadened by a Lorentzian with a width of $0.1|t|$.  Insets: The $x$ dependence of the Drude weight $D$ as well as the integrated total weight $K$.}
\end{center}
\end{figure}

Figure~\ref{OptFig} shows the dependence of the optical conductivity $\sigma(\omega)$ on the carrier concentration $x$ for the $N=20$ $t$-$t'$-$t''$-$J$ lattice.  At $x=0.1$, there is a broad-peak structure at around $\omega\sim t$ in addition to the Drude contribution centered at $\omega=0$ for both the hole and electron dopings.  Such a broad-peak structure is known to be incoherent charge excitations accompanied by magnetic excitations.~\cite{Dagotto}  This is physically characterized as an excitation from the AF ground state to an excited state where wrong spin bonds are created by the motion of carriers.  As a result of the presence of the broad peak separated from the Drude contribution, a gap-like feature, i.e., a pseudogap, emerges at around $\omega\sim 0.5t$.

For electron doping, it has been discussed~\cite{Tohyama2} that the pseudogap is very sensitive not only to $J$ but also $t'$ and $t''$: Increasing the absolute values of $t'/t$ and $t''/t$, the gap increases in energy.  Such a pseudogap feature in $\sigma(\omega)$ has been clearly observed in electron-doped NCCO.~\cite{Onose}  In Fig.~\ref{OptFig}(a), the gap feature is also seen in hole-doped case at $x=0.1$.  Although the gap feature has not been clearly reported in the normal state of hole-doped LSCO, a broad peak can be seen at $\omega\sim0.5$~eV for $x\leqslant 0.06$.~\cite{Uchida}  The calculated broad-peak structure in hole doping probably corresponds to the broad peak observed experimentally.  

At $x=0.2$, a remarkable difference appears between hole and electron dopings in Fig.~\ref{OptFig}: A pseudogap remains in electron doping accompanied by a peak at $\omega=0.7|t|$, while it disappears in hole doping.  Since such a gap feature is related to magnetic excitations as discussed above, the difference should reflect the difference of magnetic properties.   In fact, AF correlation at $x=0.2$ behaves differently as already shown in Fig.~\ref{CspinFIG}, where the AF order is expected for electron doping, while the AF correlation length is very short (smaller than 2 lattice units) in hole doping.  This clearly demonstrates the fact that charge dynamics is strongly influenced by AF spin correlation.  At $x=0.3$, $\sigma(\omega)$ in electron doping shows similar behaviors to the hole-doped case.  This is reasonable because the concentration of $x=0.3$ is enough to kill AF correlation. 

It is also interesting to compare the peak position in $\sigma(\omega)$ with the gap in the single-particle spectral function $A(\mathbf{k},\omega)$ discussed in Sec.~\ref{SpectralFunction}.  In electron doping, the values of the gap $E_\mathrm{gap}$ at around $\mathbf{k}=(\pi/2,\pi/2)$ are $1.0t$ and $0.74t$ for $x=0.1$ and $0.2$, respectively (see Fig.~\ref{Gap_pi/2_pi/2}).  These numbers almost agree to the peak positions in $\sigma(\omega)$ as shown in Fig.~\ref{OptFig}(b).  Such an agreement indicates that the pseudogap in $\sigma(\omega)$ and the gap in $A(\mathbf{k},\omega)$ in electron doping have the same origin.  Needless to say, AF spin correlation is the underlying cause of the gaps.

In the inset of Fig.~\ref{OptFig}, the Drude weight $D$ as well as the integrated total weight $K$ defined in (\ref{D}) is plotted as a function of $x$.  Both the weights increase with $x$.  Compairing the hole and electron dopings, we find that $D$ as well as $K$ is larger in electron doping than in hole doping.  Such an enhancement in electron doping, particular for $x\leqslant 0.2$, is a consequence of the interplay between the charge motion and spin background, where the AF spin background makes possible smooth sublattice-charge flows via $t'$ and $t''$.~\cite{Tohyama2}   

\subsection{Charge correlation and $d$-wave pairing}
\label{pairing}

\begin{figure}
\begin{center}
\includegraphics[width=7.5cm]{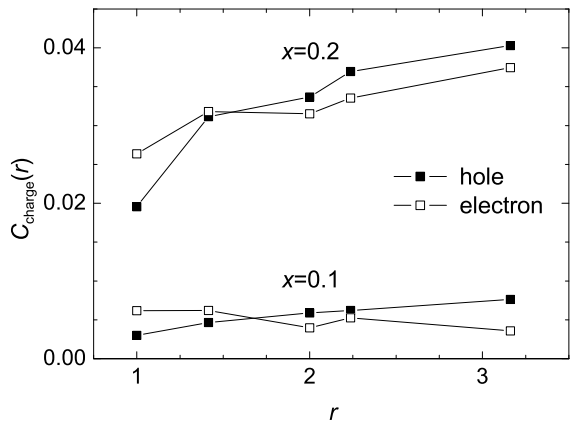}
\caption{\label{CchargeFIG}
Charge correlation $C_\mathrm{charge}(r)$ as a function of the two-carrier distance $r$ for an $N=20$ $t$-$t'$-$t''$-$J$ lattice, averaging over twisted BC.  Solid squares:  hole doping ($t=1$, $t'=-0.25$, $t''=0.12$, and $J=0.4$).  Open squares: electron doping ($t=-1$, $t'=0.25$, $t''=-0.12$, and $J=0.4$).}
\end{center}
\end{figure}

In order to consider the pairing of carriers, we first show in Fig.~\ref{CchargeFIG} the charge correlation function $C_\mathrm{charge}(r)$ that gives information on attraction between doped carriers.  Compairing hole and electron dopings, we find that the nearest-neighbor attraction ($r=1$) is stronger in electron doping than in hole doping, and vice versa for the long-distance correlation ($r=\sqrt{10}$), which  is irrespective of carrier concentration $x$.  The relatively strong nearest-neighbor attraction in electron doping is easily understood if we consider the fact that the AF order existing in electron doping is favorable for the pair formation gaining the exchange energy. For hole doping, such a force to attract two carriers is weak and thus the carriers spread over the whole system, leading to the enhancement of  long-range carrier correlation.

\begin{figure}
\begin{center}
\includegraphics[width=7.5cm]{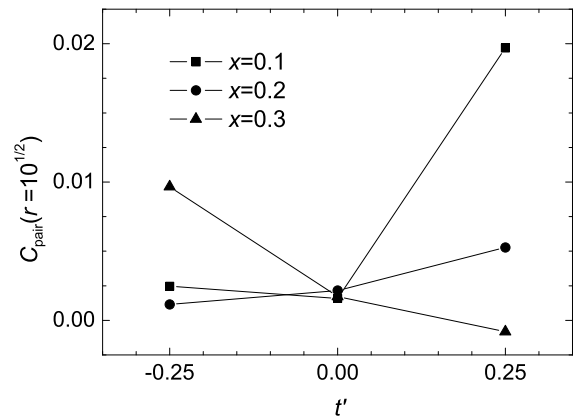}
\caption{\label{CpairFIG}
$t'$ dependence of $d_{x^2-y^2}$-wave pairing correlation $C_\mathrm{pair}(r)$ at the largest distance $r=\sqrt{10}$ for an $N=20$ $t$-$t'$-$t''$-$J$ lattice.  $t=1$ and $t''=0.12$ for $t'=-0.25$, while $t=-1$ and $t''=-0.12$ for $t'=0.25$.  $J=0.4$ for both cases.  $t'=0$ corresponds to the $t$-$J$ model.  Solid squares, circles, and triangles are for $x=0.1$, $x=0.2$, and $0.3$, respectively.}
\end{center}
\end{figure}

Since the nearest-neighbor charge attraction is strong in electron doping, short-range pairing of two electrons is also expected to be strong as compared with hole doping.   However, this does not automatically mean that superconducting pair-pair correlation is strong.  In Fig.~\ref{CpairFIG}, we show the $d$-wave pairing correlation $C_\mathrm{pair}(r)$ at the largest distance ($r=\sqrt{10}$) in the $N=20$ lattice.  We find that the $d$-wave pairing correlation is strongly enhanced for electron doping ($t'=0.25$) at $x=0.1$, which is consistent with a density-matrix renormalization-group calculation.~\cite{White}  With increasing $x$ from $0.1$ to $0.2$, the correlation decreases rapidly.  The enhancement of pairing correlation for $t'=0.25$ predominantly comes from the enhancement of pairing itself as indicated in Fig.~\ref{CchargeFIG}.  In the momentum space, the strong pairing originates from the large single-particle spectral weights near the Fermi level at $\mathbf{k}=(\pi,0)$, as shown in Fig.~\ref{Akwx=0.1ele}.  This is easily understood if we express the pairing operator $\Delta_i$ in (\ref{d}) as $2/N \sum_\mathbf{k} (\cos k_x -\cos k_y) \sum_\sigma (-1)^\sigma \tilde{c}_{\mathbf{k},\sigma}\tilde{c}_{-\mathbf{k},-\sigma}$: The $d$-wave operator has the largest amplitude at $\mathbf{k}=(\pi,0)$ and thus the large single-particle occupation at this $\mathbf{k}$ seen in electron doping at $x=0.1$ gives rise to large pairing interaction.  

The enhancement of $d$-wave pairing correlation in electron doping is accompanied by an enhancement of AF correlation.  We speculate that the AF correlation exceeds the pairing correlation near half filling, i.e., the AF order overcomes the superconducting order.  With increasing electron concentration, AF correlation weakens and finally pairing correlation may become dominant, resulting in a transition from AF to superconducting order as observed experimentally.  Such a picture may have some connections with recent data for Pr$_{1-x}$LaCe$_x$CuO$_4$ indicating the coexistence of AF order and superconductivity in the vicinity of the transition.~\cite{Fujita2}

An important point to notice further is that the AF-superconducting transition may be accompanied by a topology change of the Fermi surface from small to large ones.  At the same time, the gap at the Fermi level along the nodal direction is expected to be closed in order for $d_{x^2-y^2}$-wave superconductivity to be induced.  However, in the present calculations, the critical electron concentration $x$, where the gap closes ($x\sim0.3$), is higher than experimental values of the AF-superconducting transition ($x\sim 0.1$$-$$0.15$) or a quantum phase transition~\cite{Dagan} ($x=0.165$).  The discrepancy may indicate the presence of additional effects that have not been included in the present $t$-$t'$-$t''$-$J$ model: for instance (1) the $x$ dependence of the parameter values, which has been incorporated into the studies of a $t$-$t'$-$t''$-$U$ model,~\cite{Kusuko,Markiewicz,Kyung1,Kyung2,Senechal} and (2) the effect of inhomegeneity, the presence of which has been reported in local probes such as muon-spin relaxation~\cite{Sonier} and nuclear magnetic resonance~\cite{Zamborszky,Bakharev} experiments.  In any case, we may need to clarify the origin of the discrepancy.  This still remains as a future issue.  

In hole doping, the $d$-wave pairing correlation shows small value up to $x=0.2$ (see $t'=-0.25$ in Fig.~\ref{CpairFIG}).  However, it is enhanced at $x=0.3$.  This is consistent with a recent theoretical study,~\cite{Shih} where the enhancement is attributed to the increase of the occupation number at $\mathbf{k}=(\pi,0)$ whose position approaches the Fermi level.

\section{Summary}
\label{Summary}

In this paper, we have examined the doping dependence of magnetic and electronic properties in the hole- and electron-doped cuprates by using the exact diagonalization technique for the $t$-$t'$-$t''$-$J$ model.  In order to reduce finite-size effects in small-size lattice, twisted BC are introduced instead of standard periodic BC.  For the calculation of correlation functions, we have averaged the results for various twisted boundary conditions.  The single-particle spectral function has been obtained for all momenta in the Brillouin zone by changing the twist.

We find that the fact that AF spin correlation remains strong in electron doping in contrast to the case of hole doping, which has been obtained under the periodic BC,~\cite{Tohyama1,Tohyama2,Tohyama3} does not change even if the averaging procedure over the twist is employed.  This confirms asymmetric magnetic properties in the $t$-$t'$-$t''$-$J$ model.  This necessarily leads to a remarkable electron-hole asymmetry in the dynamical spin structure factor and two-magnon Raman scattering.  The doping dependence of these quantities in electron doping is qualitatively consistent with recent experimental data,~\cite{Fujita1,Sugai2} indicating the justification for the use of the $t$-$t'$-$t''$-$J$ model.

Using the twisted BC, we have also uncovered dramatic differences in the single-particle spectral function between hole and electron dopings.  In hole doping, the quasiparticle band for $x=0.1$ is gapless at the Fermi level along the nodal $(0,0)$-$(\pi,\pi)$ direction, but a gapped behavior emerges near the antinodal region.  The Fermi surface map shows a Fermi arc behavior, consistent with ARPES data in the underdoped cuprates.  It is important to notice that the presence of $t'$ and $t''$ is essential to the Fermi arc.  The gap near the antinodal region disappears at $x=0.2$.  In contrast, the gap appears near the nodal region in electron doping up to $x=0.2$.  The gap is found to be correlated with the strength of AF correlation, indicating that the gap is magnetically driven.  In addition to the gap, an electron pocket is clearly seen at around $(\pi,0)$ for $x=0.1$.

In the optical conductivity, a pseudogap feature clearly appears in the electron-doped system up to $x=0.2$ under the averaging procedure over the twist.  The origin of the gap is attributed to the strong AF correlation in the spin background.  In fact, we find that the pseudogap has the same magnitude as that in the spectral function along the nodal direction, confirming the same origin.

Comparing the calculated spectral function and optical conductivity with recent experimental data,~\cite{Onose,Armitage2} we find that the presence of the electron pocket and pseudogap agrees with the experimental data but only for $x\le 0.15$.  In particular, the critical electron concentration $x$ where the gap closes is $x\sim0.3$ for the present calculation, but this is higher than experimental values of the AF-superconducting transition ($x\sim 0.1$$-$$0.15$) or a quantum phase transition~\cite{Dagan} ($x=0.165$).  This discrepancy may indicate the presence of additional effects that have not been included in the present $t$-$t'$-$t''$-$J$ model.  We may need to clarify the origin of the discrepancy but leave this as a future issue.

In terms of pairing of carriers, the $d$-wave pairing correlation function is examined, and the pairing is found to be enhanced in the underdoped region of electron-doped system and also in the overdoped region of hole-doped one, consistent with previous studies under the periodic and open BC.~\cite{White,Shih}  In electron doping, AF correlation is also enhanced in the same concentration.  We thus speculate that AF correlation exceeds pairing correlation near half filling, but with increasing electron concentration AF correlation weakens and finally pairing correlation may become dominant.  Although the electronic states in the normal state of high-$T_c$ cuprates, including asymmetry between hole and electron doping, are found to be described well by the $t$-$t'$-$t''$-$J$ model, the relation of the $t$-$t'$-$t''$-$J$ model to the $d$-wave superconductivity in addition to the competition between AF order and superconductivity in electron doping remains to be resolved in the future.  

\section*{ACKNOWLEDGMENTS}
I would like to thank S. Maekawa for enlightening discussions.  I am also grateful to M. Fujita, D. H. Lu, M. Matsuda, Y. Onose,  Z.-X. Shen, S. Sugai, K. Yamada, and G.-q. Zheng for useful discussions.  This work was supported by CREST, NAREGI Nanoscience Project, and Grant-in-Aid for Scientific Research from the Ministry of Education, Culture, Sports, Science and Technology of Japan.  The numerical calculations were performed in the supercomputing facilities in ISSP, University of Tokyo and IMR, Tohoku University.

\end{document}